 \pdfoutput=1
\documentclass[twocolumn,secnumarabic,amssymb, nobibnotes,aps, prd,floatfix,superscriptaddress]{revtex4-2} 

\usepackage{makecell}

\newtheorem{algorithm}{Algorithm}
\usepackage[shortlabels]{enumitem}
\usepackage{enumitem}
\newlist{primenumerate}{enumerate}{1}
\setlist[primenumerate,1]{label={\arabic*$'$}}
\usepackage{caption}
\usepackage{subcaption}
\usepackage{dsfont}
\usepackage{hyperref}
\hypersetup{colorlinks=true, allcolors=blue}
\usepackage{amsmath}
\usepackage{xcolor}
\usepackage{comment}
\usepackage{graphicx}
\usepackage{mathtools}
\usepackage{bbm}
\usepackage{lipsum}
\usepackage{soul}

\newcommand{\bg}{\boldsymbol{g}}
\newcommand{\bS}{\boldsymbol{S}}

\newcommand{\bs}{{\boldsymbol{\sigma}}}
\newcommand{\bd}{\boldsymbol{\delta}}
\newcommand{\bt}{{\boldsymbol{\theta}}}

\newcommand{\Cov}{\mathrm{Cov}}

\newcommand{\vertiii}[1]{{\left\vert\kern-0.25ex\left\vert\kern-0.25ex\left\vert #1 
    \right\vert\kern-0.25ex\right\vert\kern-0.25ex\right\vert}}

\definecolor{gr}{RGB}{0, 158, 115}

\definecolor{vb}{RGB}{66, 120, 245}

\begin{document}

\title{Improved energies and local energies with weighted variational Monte Carlo}%

\author{Huan Zhang}%
\affiliation{Courant Institute of Mathematical Sciences, New York University, New York 10012, USA}

\author{Robert J. Webber}
\affiliation{Department of Mathematics, University of California San Diego, La Jolla, CA 92093, USA}

\author{Michael Lindsey}%
\affiliation{Department of Mathematics, University of California Berkeley, CA 94720, USA}

\author{Timothy C. Berkelbach}
\email{t.berkelbach@columbia.edu}
\affiliation{Department of Chemistry, Columbia University, New York, NY 10027, USA}
\affiliation{Initiative for Computational Catalysis, Flatiron Institute, New York, NY 10010, USA}

\author{Jonathan Weare}%
\email{weare@nyu.edu}
\affiliation{Courant Institute of Mathematical Sciences, New York University, New York 10012, USA}

\begin{abstract}
Neural network parametrizations have increasingly been used to represent the ground and excited states in variational Monte Carlo~(VMC) with promising results. 
However, traditional VMC methods only optimize the wave function in regions of peak probability.
The wave function is uncontrolled in the tails of the probability distribution, which can limit the  accuracy of the trained wavefunction approximation.
To improve the approximation accuracy in the probability tails, this paper interprets VMC as a gradient flow in the space of wave functions, followed by a projection step.
From this perspective,
arbitrary probability distributions can be used in the projection step, allowing the user to prioritize accuracy in different regions of state space. 
Motivated by this theoretical perspective, the paper tests a new weighted VMC method on the antiferromagnetic Heisenberg model for a periodic spin chain.
Compared to traditional VMC, weighted VMC reduces the error in the ground state energy by a factor of $2$ and it reduces the errors in the local energies away from the mode by large factors of $10^2$--$10^4$. 

\end{abstract}

\maketitle
\section{Introduction}
The introduction of neural network wave function parametrizations~\cite{carleo2017solving,luo2019backflow,pfau2020ab,hermann2020deep} has sparked renewed interest in the variational Monte Carlo  (VMC)~\citep{gubernatis2016quantum,becca2017quantum,PhysRevLett.121.167204, nomura2021helping,pathak2021excited,entwistle2023electronic, filippi2009absorption,zimmerman2009excited,zhao2016efficient} approach to approximating quantum many-body ground and excited states. In its simplest version, VMC approximates the ground state of a given system by minimizing an energy functional $E(\psi_\bt)$ over a parametrized class of wave functions $\psi_\bt$.
The approach uses Monte Carlo sampling 
to estimate the parameter gradient $\nabla_\bt E(\psi_\bt)$ needed for energy minimization.


Neural quantum states have shown tremendous promise within VMC, but they also lead to a problem of generalization error~\cite{westerhout2020generalization}. 
The combination of a nonlinear wave function representation with Monte Carlo approximation can lead to low accuracy wave function estimates away from regions of high probability, and it can even lead to unstable optimization \cite{webber2021rayleigh,Zhang2023understanding}. 

This paper focuses on computing the ground state of 
the antiferromagnetic Heisenberg XXX model, which has proved computationally challenging for VMC~\cite{park2020geometry, Zhang2023understanding} even though the ground state can be derived analytically using the Bethe ansatz~\cite{Karbach1998Bethe}.
Our previous work~\cite{Zhang2023understanding} documents the generalization error in the estimated ground state for the XXX model.
This error is concentrated in the region of ferromagnetic configurations far away from the mode of the wave function.
To help reduce this error, we proposed \cite{Zhang2023understanding} adding a term to the VMC objective function that prevents the wave function from growing in the ferromagnetic region. 
However, this approach does not ensure relative accuracy in the approximation of the ground state in the region of low $\lvert \psi_{\bt} \rvert^2$. 

The current paper presents a new approach to improve the wave function accuracy,
based on understanding VMC optimization as a gradient flow followed by a projection step~\cite{stokes2020quantum, neklyudov2023wasserstein}.
In this theoretical framework, the wave function is first optimized using a gradient flow that reduces the energy. Then the evolved wave function is projected onto the space of parameterized wave functions. 
The evolve-and-project approach is at the core of time-dependent variational principle commonly used to optimize wave function approximations \cite{diract1930note, PhysRevB.94.165116}.

The projection step depends on the choice of distance function,
since it identifies the element within the parametric class with the \emph{smallest distance} to the evolved wave function.
This paper introduces the idea of a weighted distance that enhances the projection accuracy in low-probability regions.
The weighted distance leads to a modified formula for the VMC parameter updates, which can be evaluated using Monte Carlo samples drawn from a weighted wave function density.
Conversely, the user is free to generate samples from any convenient distribution, and the parameter update calculated with the samples corresponds to a projection step with an associated set of weights.
This new ``weighted VMC'' approach does not require prior knowledge of the problem and should be widely applicable.

Our numerical experiments indicate the success of weighted VMC applied to the Heisenberg XXX model.
The experiments identify several sampling methods leading to improved energies and local energies throughout the configuration space.
Mixed tempering and well-tempered metadynamics~\cite{metadynamics2002,welltemperedmetadynamics2008} are broadly applicable methods that yield a factor of $2$ energy improvement compared to traditional VMC.
Additionally, uniform CV sampling is based on identifying a physically important collective variable.
It yields competitive energy estimates, and it improves the local energies in targeted regions by factors up to $10^4$.

The paper is structured as follows. 
Sec.~\ref{sec:preliminaries} provides background.
Sec.~\ref{sec:Weighted-VMC} introduces weighted VMC. Sec.~\ref{sec:examples} shows examples of weighted VMC sampling distributions. Sec.~\ref{sec:numresults} presents numerical results. Sec.~\ref{sec:conclusion} concludes.

\section{Background}\label{sec:preliminaries}

This section first introduces the XXX model and the RBM wave function ansatz.
Then, it describes the standard VMC optimization using stochastic reconfiguration.

\subsection{Model and ansatz}
This paper focuses on the antiferromagnetic Heisenberg XXX model for $N$ spin-$1\slash 2$ particles in a one-dimensional periodic chain.
After a stoquastic transformation~\cite{bishop2000marshall}, the Hamiltonian for the XXX model is
\begin{align}
\label{eq:hamiltonian}
    \hat{H} = \sum_{i=1}^{N} \bigl(
    -\hat{\sigma}_i^x \hat{\sigma}_{i+1}^x
    + \hat{\sigma}_i^y \hat{\sigma}_{i+1}^y
    + \hat{\sigma}_i^z \hat{\sigma}_{i+1}^z \bigr),
\end{align}
where $\hat{\sigma}_i^x$, $\hat{\sigma}_i^y$, and $\hat{\sigma}_i^z$ are the Pauli operators for the $i$th spin. Here and throughout, periodic boundary conditions are implied via the identification $\hat{\sigma}_{N+i}=\hat{\sigma}_i$. 
The Hamiltonian admits a unique, nonnegative-valued ground state wave function due to the Perron-Frobenius theorem.

The Hamiltonian for the XXX model commutes with the translation operator, so each ground and excited state $\psi$ must satisfy a symmetry relationship 
\begin{equation*}
    \psi (\mathcal{T}_j \hat\bs) = {\rm e}^{\mathsf{i}  2\pi kj / N} \psi (\bs),
\end{equation*}
where $T_j$ is the translation operator, $(T_j \hat\bs)_i= \hat\bs_{i-j}$, and $k \in \{0, 1, \ldots, N-1\}$ is the symmetry class.
Hereafter we focus on the $k = 0$ subspace, which contains the ground state due to the Perron-Frobenius theorem.

We define the variational wave function using a basis of simultaneous eigenstates of the operators $\{\hat{\sigma}_i\} = \{\hat{\sigma}_i^z\}$, where we drop the $z$ indicator for notational simplicity.
Using this basis, our wave function ansatz~\cite{carleo2017solving} can be written as
\begin{multline*}
\psi_{\bt}(\bs) = \\
    \sum_{(h_j)_{1 \le j \leq N}} \exp\Biggl(\sum_{i=1}^{N} a_i \sigma_i
    +
    \sum_{j=1}^{\alpha N} b_j h_j
    +
    \sum_{i=1}^N \sum_{j=1}^{\alpha N} W_{ji} h_j \sigma_i\Biggr),
\end{multline*}
where $\sigma_i \in \{-1,+1\}$ are spin variables,
$h_j \in \{-1,+1\}$ are an additional set of $\alpha N$ hidden spin variables,
and $\bt=\{\boldsymbol{a},\boldsymbol{b},\boldsymbol{W}\}$ are variational parameters.
This ansatz is an example of a restricted Boltzmann machine (RBM)~\cite{tieleman2008training}, which is a one-hidden-layer neural network that has been widely used in VMC in recent years~\cite{carleo2017solving,Zhang2023understanding,webber2021rayleigh}.

Following \cite{carleo2017solving, webber2021rayleigh}, we modify the RBM ansatz by summing over the hidden spins $h_k$ and enforcing translation symmetry: 
\begin{equation}
\label{eq:ansatz}
\psi_{\bt}(\bs) =
    \prod_{k=1}^\alpha \prod_{j=1}^N \cosh \biggl(b_k + \sum_{i=1}^N W_{ki}\sigma_{i+j} \biggr).
\end{equation}
This reduces the set of variational parameters to 
$\bt=\{\boldsymbol{b},\boldsymbol{W}\}$.
Throughout the numerical experiments, we use the width parameter $\alpha = 50$, but see Appendix~\ref{app:alpha} for further discussion.

\subsection{Variational Monte Carlo}
In this section, we adopt the standard $L^2$ inner product and norm:
\begin{equation*}
    \langle \psi , \phi \rangle = \sum_\bs \overline{\psi(\bs)} \phi(\bs),
    \quad \|\psi \|^2 = \sum_\bs |\psi(\bs)|^2.
\end{equation*}
Variational Monte Carlo (VMC) approximates the ground state by minimizing the Rayleigh quotient
\begin{equation}
\label{eq:energy}
    E(\psi)\coloneqq\frac{\langle \psi , H\psi \rangle}
    {\langle \psi , \psi \rangle}
\end{equation}
within a parametrized manifold $\psi = \psi_{\bt}$.
Stochastic reconfiguration (SR)~\cite{becca2017quantum} is a popular method to optimize traditional~\cite{Sorella2001GenLanczos,sorella2007weak} and more recent~\cite{carleo2017solving, pfau2020ab} wave functions within VMC.
SR is motivated by considering a small parameter update $\bt \leftarrow \bt + \bd$.
The parameter update changes the energy in a predictable way, according to the following linearized equation:
\begin{equation}
\label{eq:energyperturb}
    E_{\rm linear}(\psi_{\bt+\bd }) = E(\psi_{\bt}) 
    + \bd^*{\bg}+{\bg}^*\bd.
\end{equation}
Here, $\bg$ is the Wirtinger gradient~\cite{Wirtinger1927} of the energy, defined by
\begin{equation}
\label{eq:gradient}
    \bg = \frac{\langle \nabla_{\bt} \psi_\bt, [H - E(\psi)] \psi_\bt \rangle}{\langle \psi_\bt, \psi_\bt \rangle}.
\end{equation}
Each step of SR minimizes the linearized energy \eqref{eq:energyperturb} subject to quadratic penalty terms.
The first quadratic penalty is based on the square angle between wave functions~\cite{webber2021rayleigh}
\begin{equation*}
    \bd^*\bS\bd \approx (\angle(\psi_\bt, \psi_{\bt + \bd}))^2 = \biggl(\arccos \frac{\left|\langle \psi_\bt, \psi_{\bt+\bd}\rangle\right|}{\|\psi_\bt\|\|\psi_{\bt+\bd}\|}\biggr)^2,
\end{equation*}
where $\bS$ is the overlap matrix defined by
\begin{equation}
\label{eq:overlap}
    \bS = \frac{\langle \nabla_{\bt} \psi_\bt, \nabla_{\bt} \psi_\bt^\top \rangle}{\langle \psi_\bt, \psi_\bt \rangle} - \frac{\langle \nabla_{\bt} \psi_\bt, \psi_\bt \rangle}{\langle \psi_\bt, \psi_\bt \rangle} 
    \frac{\langle \psi_\bt, \nabla_{\bt} \psi_\bt^\top \rangle}{\langle \psi_\bt, \psi_\bt \rangle}.
\end{equation}
The second quadratic penalty is based on the square magnitude of the update, $\eta \lVert \bd \rVert^2$.
In summary, SR solves the optimization problem
\begin{equation*}
    \operatornamewithlimits{argmin}_{\bd}\biggl[\frac{1}{\epsilon}\bd^*(\bS+\eta\boldsymbol{I})\bd+\bd^*{\bg}+{\bg}^*\bd\biggr],
    \quad \text{for } \eta \geq 0.
\end{equation*}
The resulting parameter update is the solution to the positive definite linear system
\begin{equation}
\label{eq:exact}
    (\bS+\eta \boldsymbol{I}) \bd = \epsilon \bg.
\end{equation}

In practice, the exact SR update \eqref{eq:exact} is hard to calculate, so a Monte Carlo approximation is used.
Suppose one can sample from the wave function density
\begin{equation*}
    \rho_\bt(\bs)=\frac{|\psi_\bt(\bs)|^2}{\sum_{\bs'} |\psi_\bt(\bs^\prime)|^2}
\end{equation*}
using a population of Markov chain Monte Carlo walkers (see Sec.~\ref{sec:examples}).
Then the energy \eqref{eq:energy} can be reformulated and estimated as an expectation 
\begin{equation}\label{eq:EEloc}
    E(\psi_\bt) 
    = \sum_\bs E_\mathrm{loc}(\bs) \rho_\bt(\bs)
    = \mathbb{E}_{\rho} \left[ E_\mathrm{loc}(\bs) \right],
\end{equation}
where $E_\mathrm{loc}$ denotes the ``local energy'', defined by
\begin{equation*}
    E_\mathrm{loc}(\bs) = \frac{(H\psi_\bt)(\bs)}{\psi_\bt(\bs)}.
\end{equation*}
The gradient vector $\bg$ and overlap matrix $\bS$ can be reformulated and estimated as covariances with respect to the wave function density $\rho_\bt(\bs)$.
The resulting SR algorithm is presented as Algorithm~\ref{alg:VMC_gs}.

\begin{algorithm}[Stochastic reconfiguration] \label{alg:sr}
Choose the parameter update $\bd$ to solve
\begin{equation}
\label{eq:sr}
    (\boldsymbol{S}+\eta \mathbf{I})\bd=-\epsilon \boldsymbol{g}.
\end{equation}
Here, $\eta \geq 0$ is a nonnegative parameter that makes $\boldsymbol{S}+\eta \boldsymbol{I}$ positive definite.
The energy $E(\psi_\bt)$, gradient vector $\boldsymbol{g}$, and overlap matrix $\boldsymbol{S}$ are defined by
\begin{subequations} \label{eq:VMC}
\begin{align}
    & E(\psi_\bt) = \mathbb{E}_{\rho} \left[ E_\mathrm{loc}(\bs) \right], \\
    &\bg = \Cov_{\rho}\left[ \frac{\nabla_\bt \psi _{\bt}(\bs)}{\psi _{\bt}(\bs)}, E_\mathrm{loc}(\bs) \right], \label{eq:gVMC} \\
    & \bS =\Cov_{\rho}\left[ \frac{\nabla_\bt \psi_{\bt}(\bs)}{\psi _{\bt}(\bs)} ,\frac{\nabla_\bt  \psi _{\bt}(\bs)}{\psi _{\bt}(\bs)} \right],\label{eq:SVMC}
\end{align}
\end{subequations}
where $\mathbb{E}_{\rho}$ and $\Cov_{\rho}$ indicate the expectation value and covariance with respect to the wave function density $\rho_\bt(\bs)$.
\label{alg:VMC_gs}
\end{algorithm} 

\section{Weighted variational Monte Carlo}
\label{sec:Weighted-VMC}

This section derives stochastic reconfiguration from an alternative perspective of gradient flows, building on the approach in \cite{neklyudov2023wasserstein}.
Then it proposes a modified gradient flow method to improve generalization error in VMC.

\subsection{Gradient flow perspective}

The energy functional $E(\psi)$ has a Wirtinger gradient~\cite{Wirtinger1927} that is a $n$-dimensional vector
\begin{equation*}
    \nabla_\psi E(\psi)
    = \frac{H \psi - E(\psi) \psi}{\langle \psi, \psi \rangle}.
\end{equation*}
Therefore, a method to find the exact ground state is based on running the continuous-time gradient flow
\begin{equation}
\label{eq:dynamics}
    \partial_t \psi_t = -\nabla_{\psi_t} E(\psi_t)
    = \frac{E(\psi_t) \psi_t - H \psi_t}{\langle \psi_t, \psi_t \rangle}.
\end{equation}
Yet, this continuous-time evolution cannot be implemented exactly when the wave function lies on a parametrized manifold $\psi_t = \psi_{\bt}$.
The gradient flow pushes $\psi_{t + \epsilon}$ off the manifold, so that no parameter vector $\bt_{t+\epsilon}$ exactly corresponds to $\psi_{t+\epsilon}$ defined by Eq.~\eqref{eq:dynamics}. 

To apply gradient flow within a parametrized manifold, it is necessary to combine evolution with projection.
We first evolve the wave function from $\psi$ to $\psi_{t+\epsilon}$.
We then identify a parameter vector $\bt_{t+\epsilon}$ that projects $\psi_{t+\epsilon}$ onto the manifold:
\begin{equation}
\label{eq:projection}
    \bt_{t+\epsilon} = \operatornamewithlimits{argmin}_\bt d(\psi_{t + \epsilon}, \psi_{\bt}).
\end{equation}
The projection step can be carried out in any metric $d(\cdot, \cdot)$, and one natural choice is the Fubini-Study metric~\cite{stokes2020quantum}
\begin{equation*}
    d(\phi,\psi) = \angle(\phi, \psi) = \arccos \frac{\left|\langle \phi, \psi\rangle\right|}{\|\phi\|\|\psi\|},
\end{equation*}
which measures the angle between two unnormalized wave functions.
In the limit of small time steps $\epsilon \rightarrow 0$, the evolve-and-project method leads to a parameter update
\begin{equation*}
    \bt_{t+\epsilon} = \bt_t - \frac{\epsilon}{\lVert \psi_{\bt} \rVert^2} \bS^{-1} \bg  + \mathcal{O}(\epsilon^2).
\end{equation*}
This update corresponds to the SR method \eqref{eq:sr} with $\eta = 0$
and a modified step size
$\epsilon / \lVert \psi_{\bt} \rVert^2$.

\subsection{Weighted variational Monte Carlo}

We now exploit the flexibility afforded by the evolve-and-project method.
Given any positive weight function $\mu(\bs)$, we can 
introduce a weighted Fubini-Study metric
\begin{equation*}
    d_{\mu}(\phi,\psi) = \arccos \frac{|\langle \phi, \psi \rangle_{\mu}|}{\|\phi\|_{\mu} \|\psi\|_{\mu}},
\end{equation*}
with an inner product and norm defined by
\begin{align*}
    \langle \phi,\psi\rangle_\mu &= \sum_{\bs} \mu(\bs)\overline{\phi(\bs)}\psi(\bs), \\
    \| \psi\|^2_\mu &= \sum_{\bs} \mu(\bs) |\psi(\bs)|^2.
\end{align*}
Like the standard Fubini-Study metric, $d_\mu$ is nonnegative and $d_\mu (\phi,\psi)=0$ only when $\phi= c\psi$ where $c$ is a complex number.
However, the weight function $\mu(\bs)$ is a new feature in this approach.

The change in metric leads to a change in the projection step.
Now, as $\epsilon \rightarrow 0$, the evolve-and-project method leads to
\begin{equation*}
    \bt_{t+\epsilon} = \bt_t - \frac{\epsilon}{\|\psi_{\bt}\|_\mu^2} \tilde{\bS}^{-1} \tilde{\bg}  + \mathcal{O}(\epsilon^2).
\end{equation*}
Here, the gradient vector $\bg$ in \eqref{eq:gradient} and overlap matrix $\bS$ in \eqref{eq:overlap} are modified to become:
\begin{align*}
    \tilde{\bg} &= \frac{\langle \nabla_{\bt} \psi_\bt, (H - E(\psi)) \psi_\bt \rangle_{\mu}}{\langle \psi_\bt, \psi_\bt \rangle_{\mu}}, \\
    \tilde{\bS} &= \frac{\langle \nabla_{\bt} \psi_\bt, \nabla_{\bt} \psi_\bt^\top \rangle_{\mu}}{\langle \psi_\bt, \psi_\bt \rangle_{\mu}} - \frac{\langle \nabla_{\bt} \psi_\bt, \psi_\bt \rangle_{\mu}}{\langle \psi_\bt, \psi_\bt \rangle_{\mu}} 
    \frac{\langle \psi_\bt, \nabla_{\bt} \psi_\bt^\top \rangle_{\mu}}{\langle \psi_\bt, \psi_\bt \rangle_{\mu}}.
\end{align*}
The major difference is that the $L^2$ inner products have been replaced by $\mu$-weighted inner products.
This evolve-and-project approach can be understood as an example of the time-dependent variational principle~\cite{diract1930note, PhysRevB.94.165116,Bruna_2024,zhang2024sequentialintime}.

In weighted VMC, the parameter update $\tilde{\bS}^{-1} \tilde{\bg}$ is hard to calculate exactly.
Instead, $\tilde{\bg}$ and $\tilde{\bS}$ can be estimated using covariances with respect to a weighted wave function density
\begin{align}\label{eq:weighted_rho}
    \tilde{\rho}_\bt(\bs) = \frac{\mu (\bs) |\psi_\bt (\bs)|^2 }{\sum_{\bs'} \mu (\bs') |\psi_\bt (\bs')|^2 }.
\end{align}
The new ``weighted VMC'' method based on \eqref{eq:weighted_rho} is presented as Algorithm~\ref{alg:weighted}.
As usual, the practical algorithm includes a small factor $\eta \mathbf{I}$, which helps stabilize the updates.
\begin{algorithm}[Weighted VMC] \label{alg:weighted}
Choose the parameter update $\bd$ to solve
\begin{equation}
\label{eq:weighted_update}
    (\tilde{\boldsymbol{S}}+\eta \mathbf{I})\bd=-\epsilon \tilde{\boldsymbol{g}}.
\end{equation}
Here, $\eta \geq 0$ is a nonnegative parameter that makes $\tilde{\boldsymbol{S}}+\eta \boldsymbol{I}$ positive definite.
The energy $E(\psi_\bt)$, gradient vector $\tilde{\boldsymbol{g}}$, and overlap matrix $\tilde{\boldsymbol{S}}$ are defined by
\begin{subequations} \label{eq:GPVMC_proj}
\begin{align}
    & E(\psi_\bt) = \mathbb{E}_{\rho} \left[E_\mathrm{loc}(\bs)\right], \\
    & \tilde{\bg} = \Cov_{\tilde{\rho}} \left[\frac{\nabla_{\bt} \psi _{\bt}(\bs)}{\psi _{\bt}(\bs)}, E_\mathrm{loc}(\bs) \right], \label{eq:b} \\
    & \tilde{\bS} = \Cov_{\tilde{\rho}}\left[ \frac{\nabla_{\bt} \psi_{\bt}(\bs)}{\psi _{\bt}(\bs)} ,\frac{\nabla_{\bt} \psi _{\bt}(\bs)}{\psi _{\bt}(\bs)} \right]\,,\label{eq:c} 
\end{align}
\end{subequations}
where $\mathbb{E}_{\rho}$ indicates the expectation value with respect to the wave function density $\rho_\bt(\bs)$ and $\Cov_{\tilde{\rho}}$ indicates the covariance with respect to the weighted wave function density ${\tilde{\rho}}_{\bt}(\bs)$.
\label{alg:VMC_cv}
\end{algorithm} 

Weighted VMC uses a sequence of sampling distributions that are defined iteratively.
At each optimization step, we generate samples from a weighted distribution $\tilde{\rho}_{\bt}(\bs)$ to optimize the parameter vector $\bt$.
Then, after all the optimization steps are complete, we generate samples from $\rho_{\bt}(\bs)$ to evaluate the energy $E(\psi_\bt)$.
This approach has the same computational cost as standard VMC (Alg.~\ref{alg:sr}).


What opportunities does weighted VMC provide us? 
In standard VMC, the projection step hardly emphasizes the regions of small wave function density $\rho_\bt(\bs)$. 
This lack of emphasis is reflected in Eq~\eqref{eq:VMC}, where samples from $\rho_\bt$ are used to compute the covariances: these samples overwhelmingly come from the high-probability regions.
In contrast, weighted VMC controls the relative importance of different regions using a weighted sampling distribution $\tilde{\rho}_{\bt}(\bs)$.
Increasing the relative density of $\tilde{\rho}_\bt(\bs)$ at a location $\bs$ results in greater agreement between $\psi_{t + \epsilon}(\bs)$ and the parametrized update $\psi_{\bt + \bd}(\bs)$, which can lead to higher quality at the selected location.
By adjusting the sampling distribution, we can selectively refine the wave function where refinement is needed the most.

\section{Weighted distributions} \label{sec:examples}

The key to weighted VMC is the freedom to choose a weighted distribution that emphasizes important regions of configuration space.
This section describes several distributions that can be used for sampling from 
the XXX Heisenberg model.
The section first describes the standard Markov chain Monte Carlo (MCMC) approach to sample from the unweighted wave function density. 
Then it proposes three modifications to the sampling approach that emphasize low-probability regions.

\subsection{Markov chain Monte Carlo}

Here we explain how to sample from the wave function density $\rho_{\bt}(\bs)$ using MCMC.
All the sampling methods for weighted distributions $\tilde{\rho}_{\bt}(\bs)$ are modified from this basic MCMC sampling scheme.

Our approach uses a population of $M = 100$ independent MCMC walkers, which are labeled $\bs_t^m$ for $m = 1, \ldots, M$ and $t = 0, 1, \ldots$.
We evolve each walker by proposing a new configuration $\bs_{t+1}^m$ where a randomly chosen positive spin $(\bs_t^m)_i = +1$ and a randomly chosen negative spin $(\bs_t^m)_j = -1$ are simultaneously flipped,
\begin{equation*}
     (\bs_{t+1}^m)_{\ell} = \begin{cases}
     -(\bs_t^m)_{\ell}, & \ell \in \{i, j\}, \\
         (\bs_t^m)_{\ell}, & \text{otherwise}.
     \end{cases}
\end{equation*}
We accept the ``swap move'' with probability
\begin{equation*}
    \min\biggl\{1, \frac{|\psi_\bt (\bs_{t+1}^m)|^2}{|\psi_\bt (\bs_{t}^m)|^2} \biggr\}.
\end{equation*}
Otherwise we reject and set $\bs_{t+1}^m = \bs_t^m$ instead.
We carry out 2000 Metropolis steps in this way, and we subsample the resulting $\bs$ values once per 100 steps, reducing the computation and storage costs to $20$ samples per walker.
This MCMC sampling scheme preserves the total magnetization, $\sum_{\ell=1}^n (\bs_t^m)_\ell = 0$, which is appropriate since the exact ground state of the XXX model \eqref{eq:hamiltonian} is supported on states of total magnetization zero whenever there is a bipartite lattice \cite{bishop2000marshall}.

\subsection{Mixed tempering}\label{sec:tempering}

The first version of weighted VMC uses a ``mixed tempering'' (MT) distribution, where one half of the MCMC walkers are sampled from $\rho_\bt(\bs)$ and the other half are sampled from a tempered distribution
\begin{equation*}
    \frac{\rho_\bt(\bs)^{\beta}}{\sum_{\bs}\rho_\bt(\bs)^{\beta}},
\end{equation*}
determined by an ``inverse temperature'' parameter $\beta \in [0, 1]$.
Therefore, the MT distribution can be written as
\begin{equation} 
\label{eq:mutempered}
    \tilde{\rho}_\bt(\bs) = \frac{\rho_\bt(\bs)}{2}
    +   \frac{\rho_\bt(\bs)^{\beta}}{2\sum_{\bs}\rho_\bt(\bs)^{\beta}}.
\end{equation}

\begin{figure}
    \centering
    \includegraphics[width=1.0\columnwidth]{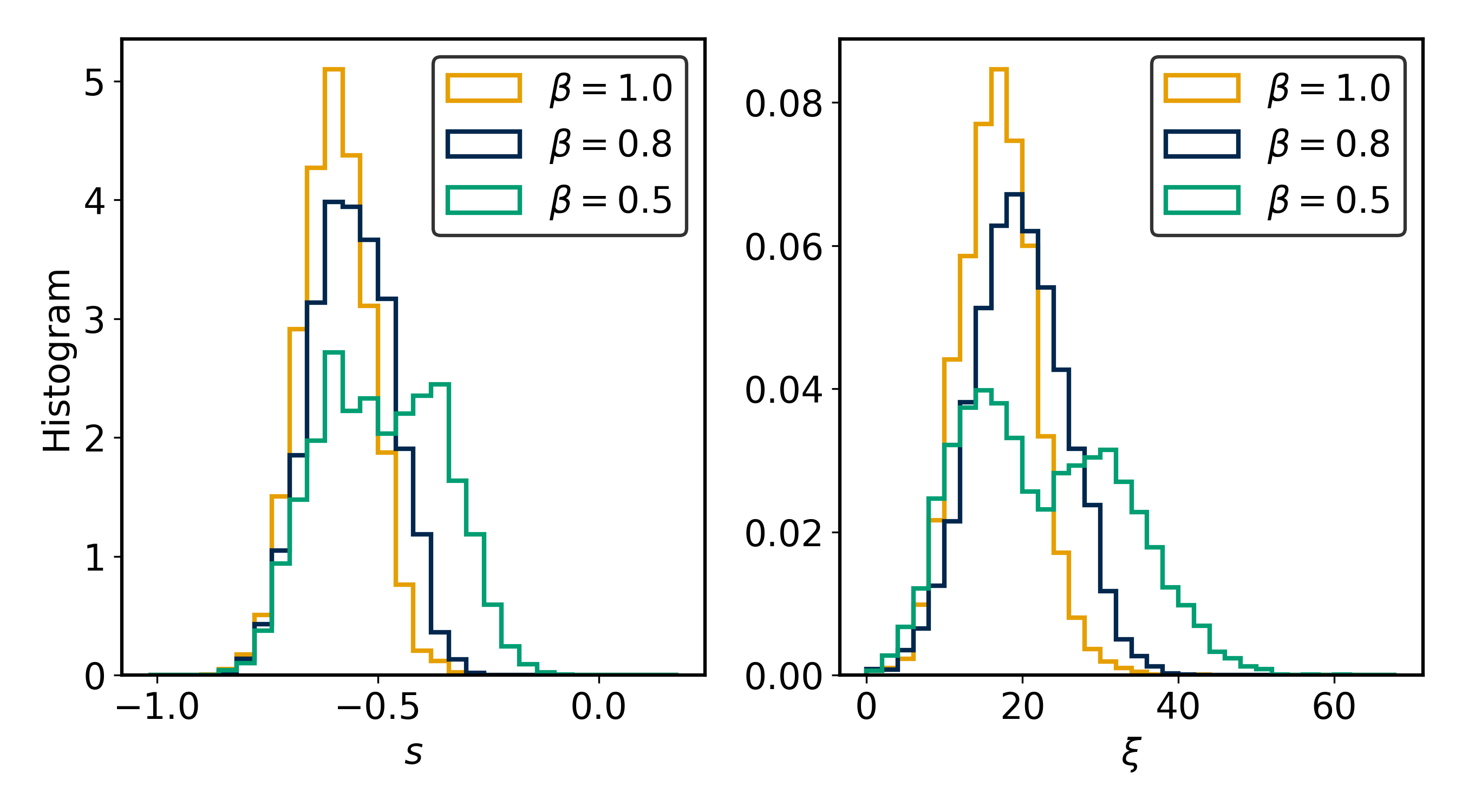}
    \caption{Histograms of the collective variables $s(\bs)$ defined in~\eqref{eq:spin-correlation} (left) and $\xi(\bs)$ defined in~\eqref{eq:Slogdensity} (right), when $4000$ MCMC samples are generated from the MT distribution with $\beta = 1.0$, $0.8$, or $0.5$.}
    \label{fig:tempdist}
\end{figure}

Fig.~\ref{fig:tempdist} visualizes the MT distribution with $\beta = 1.0$, $0.8$, and $0.5$ along two choices of ``collective variables'' (CVs).
The first CV is the nearest-neighbor spin correlation function
\begin{align}\label{eq:spin-correlation}
   s(\boldsymbol{\sigma})\equiv \frac{1}{N} \sum_{i=1}^{N} \sigma_i \sigma_{i+1}.
\end{align}
The second CV is the negative logarithm of the unnormalized wave function density
\begin{align}\label{eq:Slogdensity}
    \xi(\bs)\equiv-\log |\psi_\bt (\bs)|^2\,.
\end{align}
When $\beta=1$ in Eq.~\eqref{eq:mutempered}, the MT distribution reduces to the wave function density $\rho_\bt(\bs)$.
However, as $\beta\to 0$, the second group of walkers spreads more uniformly over the state space $\{\bs\}$.
This spreading out is reflected in the histograms in Fig.~\ref{fig:tempdist}.

\subsection{Well-tempered metadynamics}
\label{sec:WTMD}

Another version of weighted VMC is based on well-tempered metadynamics (WTMD)~\cite{metadynamics2002,welltemperedmetadynamics2008}.
WTMD is a Monte Carlo method that estimates the marginal distribution of a high-dimensional probability distribution along a user-specified CV.
WTMD drives the walkers to explore regions where the estimated marginal density is small, and it uses reweighting to recover statistics.

In this study, we apply a sampling method inspired by WTMD where the CV is the negative logarithm of the unnormalized wave function density, $\xi(\bs) = -\log |\psi_{\bt}(\bs)|^2$ as in~\eqref{eq:Slogdensity}. 
Our specific sampling procedure follows:

\begin{enumerate}
    \item \emph{Initialization.} Set $w$ and $\Delta T$ parameters for the marginal density estimation. 
    Set the lower bound $L_\xi$ and bin interval $\Delta \xi$ for the histogram of $\xi$.
    Set the initial histogram values $V = (0, \ldots, 0) \in \mathbb{R}^{n+1}$, and choose an initial configuration of walkers $\bs_0^1, \ldots,\bs_0^M$. 
    \item \emph{Sampling.} For each $t=0, \ldots, T$ and $m = 1, \ldots, M$, perform the following steps.
    \begin{enumerate}
        \item Propose an evolution from $\bs_t^m$ to $\bs_{t+1}^m$ using a random swap move.
        Accept $\bs_{t+1}^m$ with probability
        \begin{align}
        \label{eq:accept}
            && \min\biggl\{1, \frac{{\rm e}^{V(\bs_t^m)}\,|\psi_\bt (\bs_{t+1}^m)|^2}{{\rm e}^{V(\bs_{t+1}^m)}\,|\psi_\bt (\bs_{t}^m)|^2}\biggr\}.
        \end{align}
        Otherwise reject and set $\bs_{t+1}^m = \bs_t^m$.
        \item Compute a bin index $i \in \mathbb{Z}$ such that
        \begin{align*}
        && \xi(\bs_{t+1}^m)\in  (L_\xi + (i-1)\Delta \xi, L_\xi + i \Delta \xi].
        \end{align*}
        Round up to $i = 0$ if $i$ is negative.
        Round down to $i = n + 1$ if $i$ exceeds $n + 1$.
        Then update the associated histogram value
        \begin{align*}
        && V(i)=V(i) + w\, {\rm e}^{-V(i)/\Delta T}.
        \end{align*}
    \end{enumerate}
\end{enumerate}
In the WTMD sampling procedure, the walkers are driven to explore new regions because of the term ${\rm e}^{V(\bs_{t}^m)-V(\bs_{t+1}^m)}$ in the acceptance probability \eqref{eq:accept}.
After $T$ steps, the walkers are done exploring and we define the sampling distribution as the average of delta functions on the walkers' locations
\begin{equation*}
    \tilde{\rho}_{\bt}(\bs) = \frac{1}{MT} \sum_{m=1}^M \sum_{t=1}^T \delta_{\bs_t^m}(\bs).
\end{equation*}
We start fresh by re-initializing the marginal density approximation and the sampling distribution after updating the VMC parameter weights.

\subsection{Uniform CV sampling}
\label{sec:uniform_cv}

The MT and WTMD sampling distributions can be applied to any quantum system, since they only depend on the unnormalized wave function density $|\psi_\bt(\bs)|^2$.
However, a different approach leverages prior knowledge of the physical system in the form of a well-chosen CV.
For example, the ground state of the XXX model is primarily supported on configurations in the antiferromagnetic region where
\begin{equation*}
    s(\bs) = \frac{1}{N} \sum_{i=1}^{N} \sigma_i \sigma_{i+1} < 0.
\end{equation*}
In contrast, the wave function density puts little mass on the ferromagnetic region, $\{\bs\,|\, s(\bs) > 0\}$.
We can help address this imbalance
by applying weighted VMC with a target distribution
\begin{equation} 
\label{eq:mumixture} 
    \tilde{\rho}_\bt(\bs) = 0.05\, p(\bs)+ 0.95\,\rho_\bt(\bs),
\end{equation}
where $p(\bs)$ has a nearly uniform marginal distribution along the CV $s(\bs)$.
In practice, we can pre-generate samples from $p(\bs)$ and randomly drawn from the pool in each iteration of weighted VMC.
See Appendix~\ref{app:visualize} for details.

\section{Numerical Experiments}
\label{sec:numresults}

This section describes numerical experiments comparing traditional and weighted VMC methods (Algs.~\ref{alg:sr} and \ref{alg:weighted}).
We tested both VMC approaches on the XXX Heisenberg model for a periodic chain of $N = 100$ spins.
We evaluated performance based on two metrics: energy error and local energy error.

The energy error is defined as
\begin{equation}
\label{eq:energy_error}
    [E(\psi_\bt) - E_0] / N.
\end{equation}
Here, $E(\psi_\bt)$ is the energy associated with the VMC wave function $\psi_\bt$, and $E_0$ is the exact ground state energy.
Throughout this section, energies are normalized by the number of spins $N$.

The local energy error is defined as
\begin{equation}
\label{eq:local_error}
    |E_{\rm loc}(\bs) - E_0| / N.
\end{equation}
Here, $E_\mathrm{loc}(\bs) = (H\psi)(\bs) / \psi(\bs)$ is the ``local energy'', which satisfies $E_\mathrm{loc}(\bs) = E_0$ when $\psi$ is the exact ground state.
Therefore, the local energy error quantifies local inaccuracies in the VMC wave function approximation.

The energy and local energy errors were numerically evaluated as follows:
\begin{itemize}
    \item We calculated the ground state energy $E_0$ exactly, using the Bethe ansatz~\cite{Karbach1998Bethe}.
    \item To evaluate the energy $E(\psi_\bt)$ via Eq.~\ref{eq:EEloc}, we extracted the parameter vector $\bt$ after $5000$ iterations of traditional or weighted VMC.
    We ran $N = 100$ MCMC walkers for $20000$ Metropolis steps targeting the wave function density $\rho_\bt$.
    We subsampled every 100 steps and removed the first half of the time series to account for initialization bias.
    \item Last, we evaluated the local energy error for traditional and weighted VMC using a diverse set of samples described in Appendix~\ref{app:visualize}.
\end{itemize}

The numerical experiments lead to three main results:
\begin{enumerate}  
    \item Weighted VMC methods reduces the energy error by a factor of $2$, compared to traditional VMC.
    See Sec.~\ref{sec:examples}.
    \item 
    Weighted VMC methods reduce the local energy errors in the tails of the probability distribution by factors of $10^2$--$10^4$, compared to traditional VMC.
    See Sec.~\ref{sec:numresults-local_energy_away_from_the_mode}.
    \item Weighted VMC methods methods extend to higher-dimensional systems with $N = 200$ spins, yielding the same pattern of improved energies and local energies.
    See Sec.~\ref{sec:higher}.
\end{enumerate}
The rest of the section describes these results in detail.





\subsection{Improved energies}\label{sec:numresults-energyacc}
 
\begin{table}
    \centering
    \begin{tabular}{p{.25\linewidth}|p{.25\linewidth}|p{.25\linewidth}}
      Method   & Mean & Median \\
      \hline
       VMC  & 2.66$\times 10^{-6}$&2.34$\times 10^{-6}$ \\ 
       MT & 1.46$\times 10^{-6}$&1.27$\times 10^{-6}$ \\
       WTMD  &1.45$\times 10^{-6}$ &1.48$\times 10^{-6}$ \\
    \end{tabular}
    \caption{Energy errors of traditional VMC versus weighted VMC with the MT ($\beta = 0.5$) or WTMD distribution, computed from ten independent trials.}
    \label{tab:energy_error}
\end{table}

\begin{figure}
    \centering
    \includegraphics[width=1.0\columnwidth]{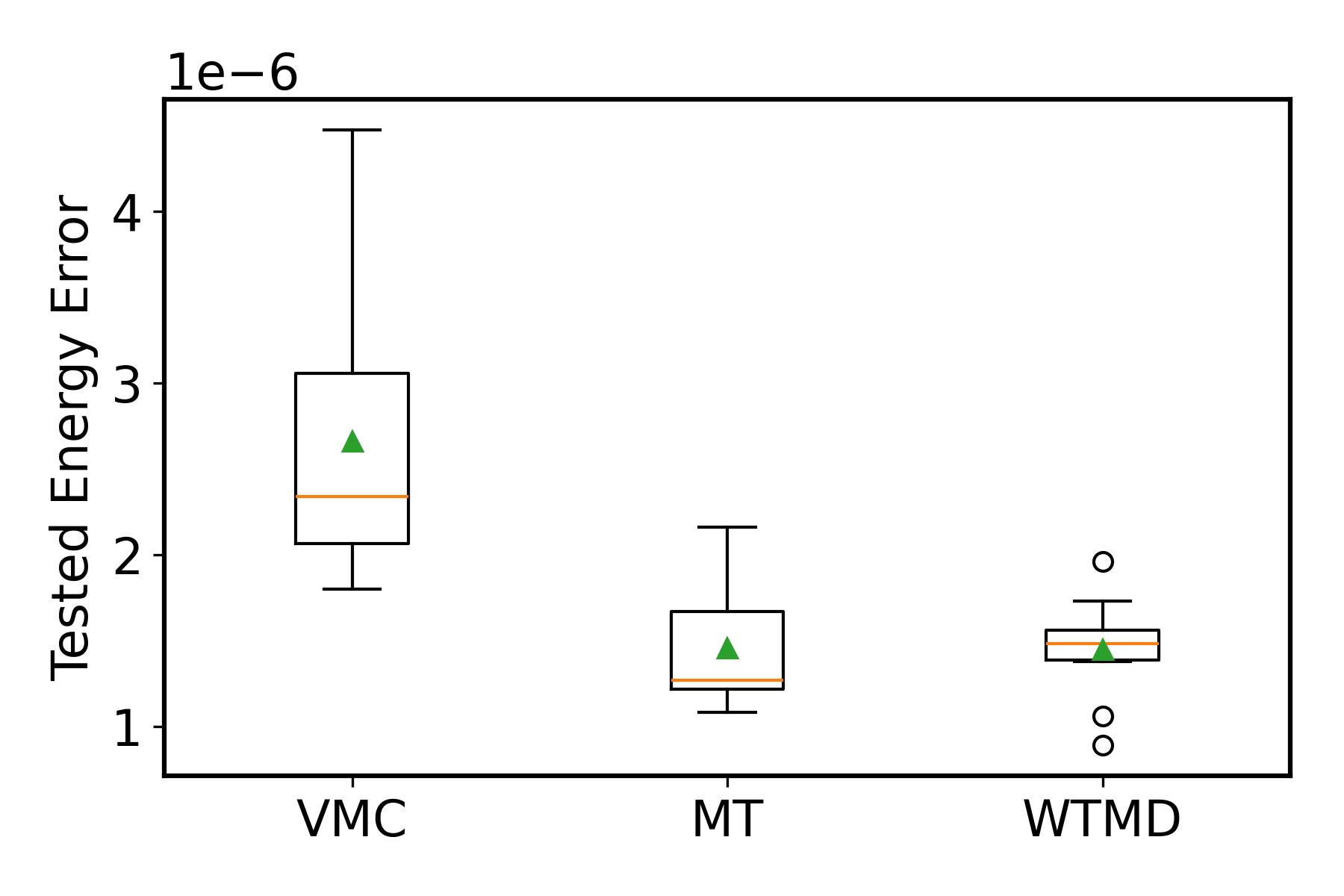}
        \caption{Boxplots of energy errors in traditional VMC (VMC) versus weighted VMC (MT, WTMD). The orange lines are median values, the green triangles are mean values.
        The dots are outliers beyond $4 \times$ the interquartile range.
        }
    \label{fig:compare_energies}
\end{figure}

As the first main result, Tab.~\ref{tab:energy_error} 
demonstrates a factor-of-two energy improvement when switching from traditional to weighted VMC using either the MT ($\beta = 0.5$) or WTMD sampling distribution (Secs.~\ref{sec:tempering} and \ref{sec:WTMD}).
The same data is visualized in a different way using boxplots in Fig.~\ref{fig:compare_energies}.
The comparison between traditional and weighted VMC is reasonable since these methods have the same computational cost.

\begin{figure}
    \centering
    \includegraphics[width=1.0\columnwidth]{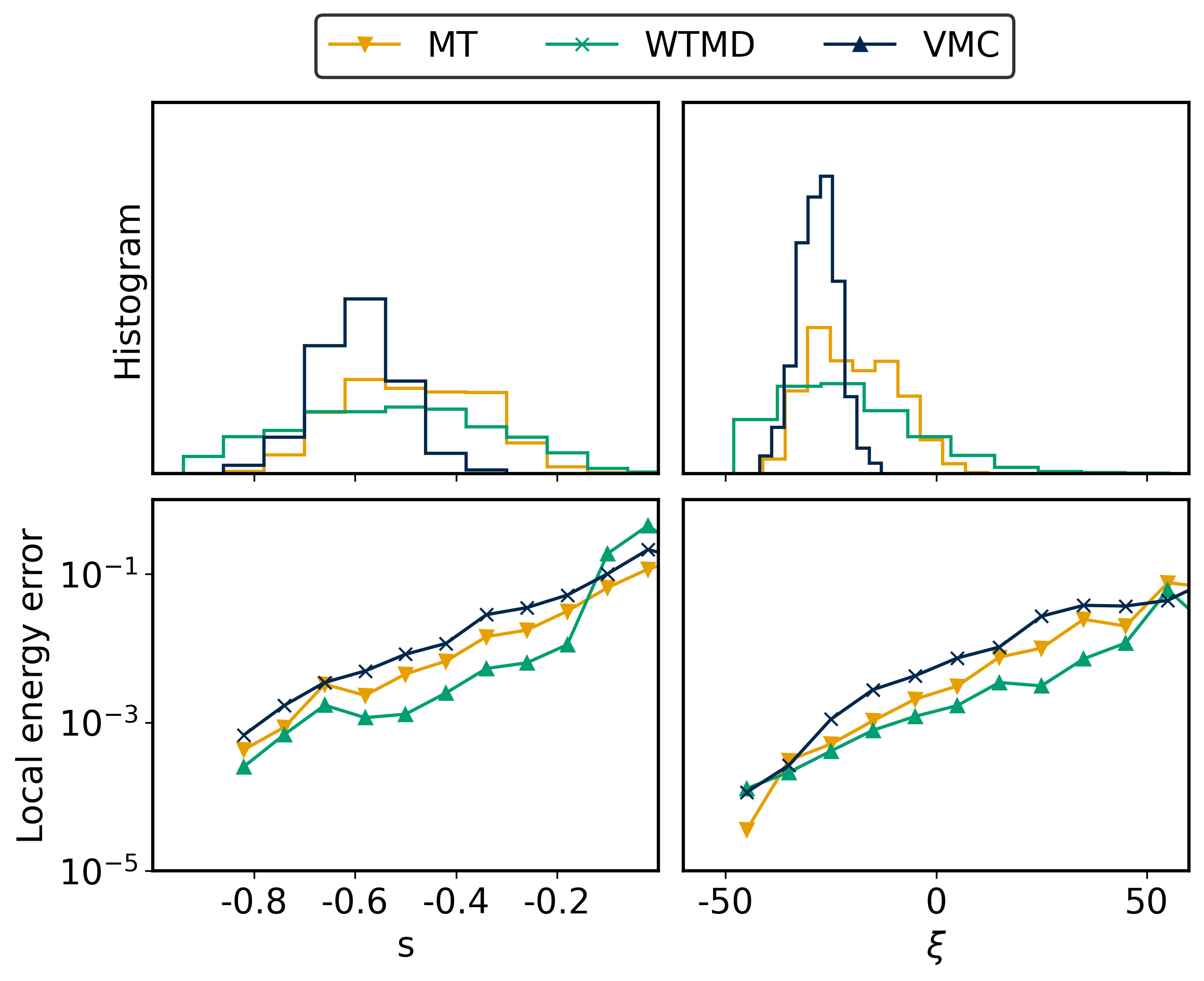}
        \caption{Histograms and magnitudes of local energy errors evaluated across two collective variables: $s(\bs)$ (left) and $\xi(\bs)$ (right).
        Curves show the average local energy error for each bin.}
    \label{fig:local_energy}
\end{figure}
 
To investigate the source of the energy improvement, we selected one run of VMC, MT and WTMD 
for further analysis.
In Fig.~\ref{fig:local_energy}, we plot histograms and the local energy errors across two
collective variables $s(\bs)$ defined in~\eqref{eq:spin-correlation} and $\xi(\bs)$ defined in~\eqref{eq:Slogdensity}.
From the histograms, we see that weighted VMC increases the number of samples in regions far away from the mode of $\rho_\bt$.
From the local energies, weighted VMC leads to improved wave function accuracy for all the CV values where the histograms are populated.
This robust improvement in the wave function quality implies a surprising conclusion: improved sampling in the region \textit{surrounding} the mode can enhance local accuracy \textit{at} the mode. 
 
We observe minor differences between MT and WTMD sampling distributions. 
The variance of the WTMD energy estimate is slightly lower than the
the variance of the MT energy estimate,
as seen in  Fig.~\ref{fig:compare_energies}.
Nevertheless, MT and WTMD lead to similar mean and median energy errors, as reported in Tab.~\ref{tab:energy_error}.
From this comparison, weighted VMC can perform well with multiple sampling distributions, and neither MT nor WTMD uniformly improves on the other.

\begin{figure}
    \centering
    \includegraphics[width=1.0\columnwidth]{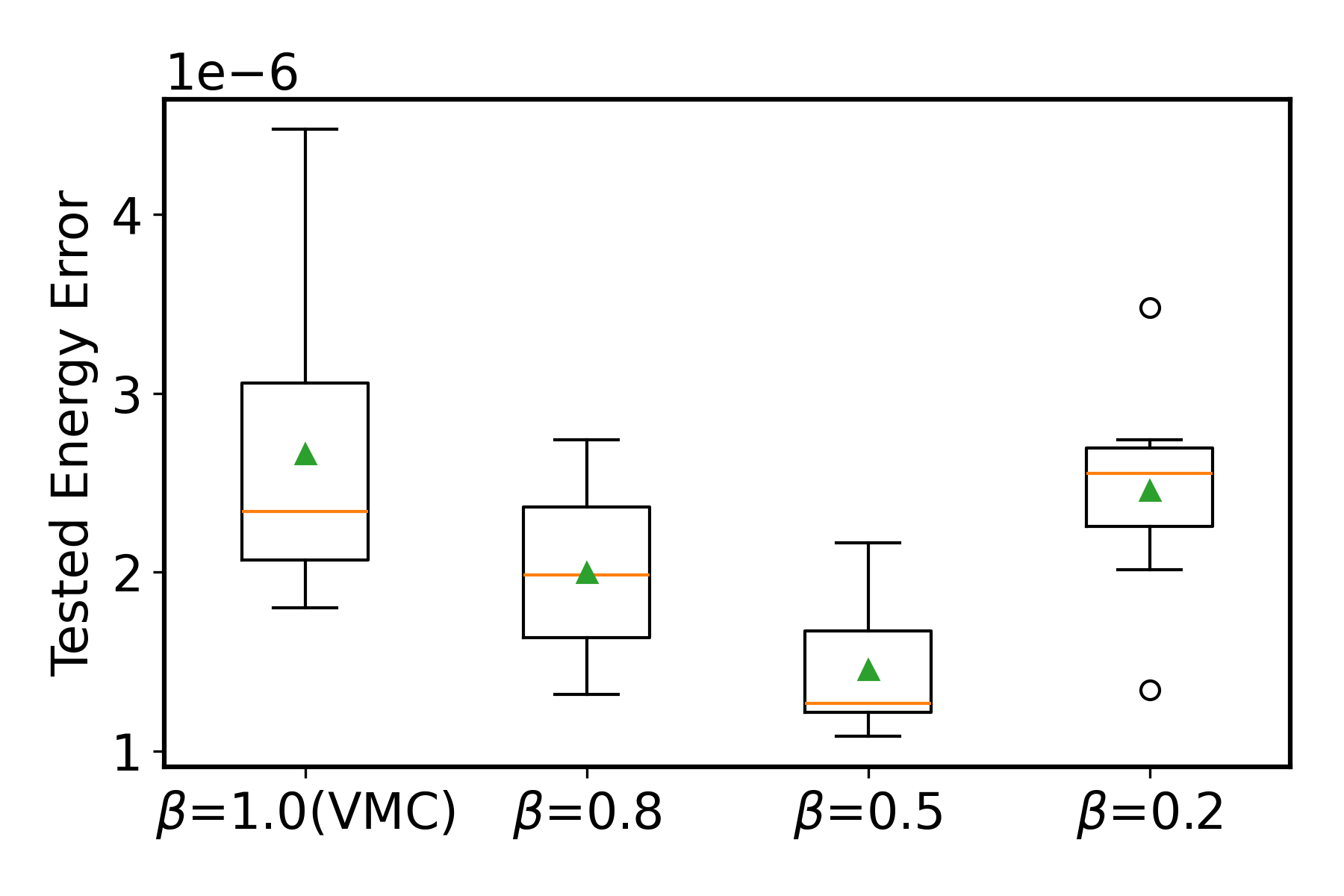}
        \caption{Boxplots of energy errors using the MT sampling distribution with parameters $\beta=1.0,0.8,0.5,0.2$.}
    \label{fig:compare_energies_temper}
\end{figure}

We also explored how the inverse temperature $\beta$ affects the energy estimate using MT sampling. Fig.~\ref{fig:compare_energies_temper} shows box plots of the MT energy errors with parameters $\beta=1.0$, $0.8$, $0.5$, and $0.2$.
The choice
$\beta=1.0$ corresponds to traditional VMC sampling.
As $\beta$ decreases, the energy error initially improves, reaching its lowest value at $\beta=0.5$. However, further reduction in $\beta$ leads to a worse energy estimate, as the samples become too dispersed to effectively refine the wave function estimate near the mode of $\rho_\bt$.
See Fig.~\ref{fig:tempdist} for an illustration of the MT distribution with several choices of $\beta$.

\subsection{Improved local energies in the tails}\label{sec:numresults-local_energy_away_from_the_mode}
Weighted VMC makes it possible to improve the local energy accuracy in low-probability regions while still maintaining an advantage over VMC for energy estimation. 
These benefits are evident when the user applies a general-purpose sampling distribution such as MT or WTMD.
The benefits are amplified when weighted VMC is applied with a physically-informed sampling distribution that drives walkers toward important tail regions identified by a user-chosen collective variable (Sec.~\ref{sec:uniform_cv}).

\begin{figure}
    \centering
 \includegraphics[width=1.0\columnwidth]{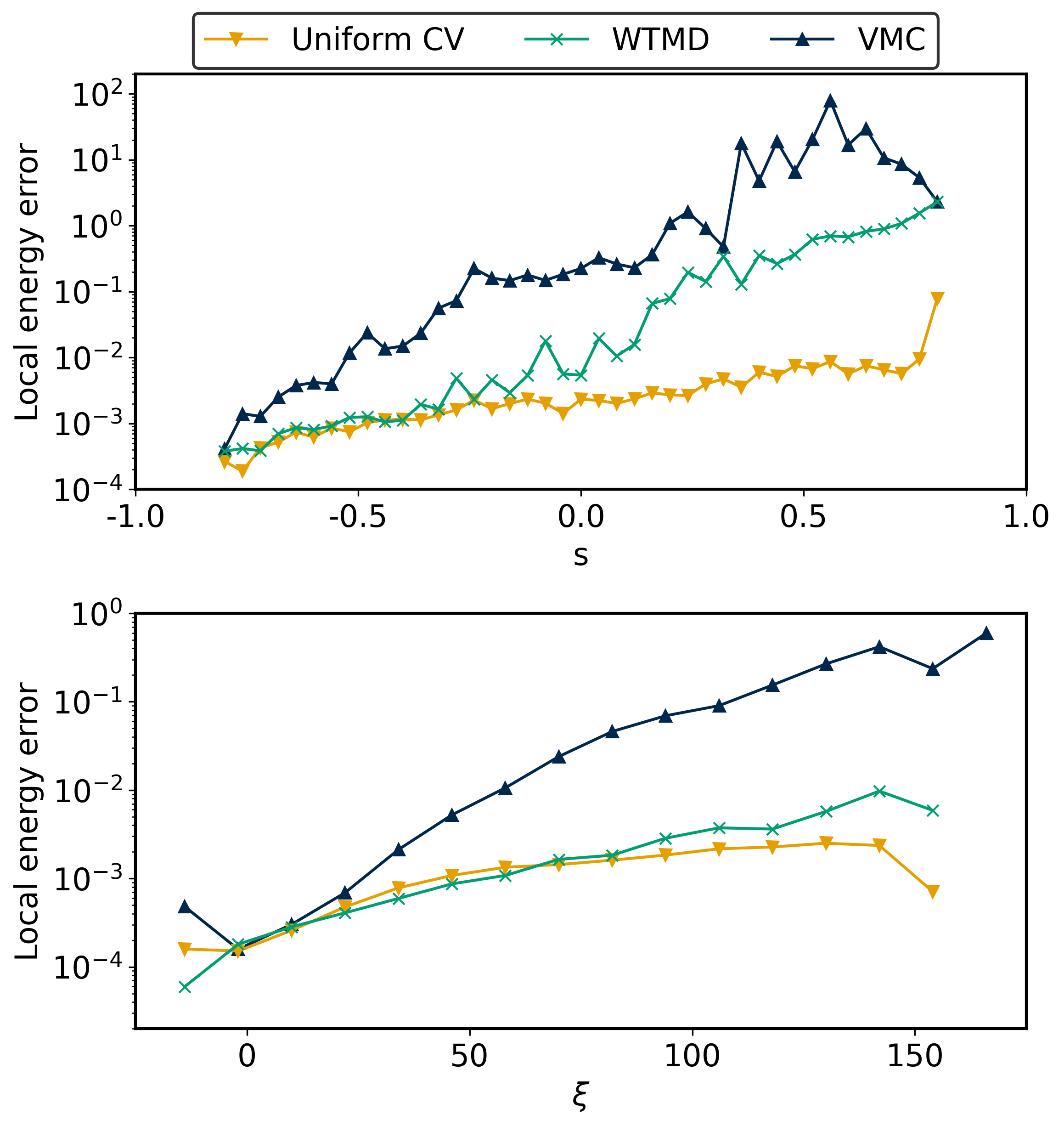}
        \caption{Local energy errors for uniform CV sampling (Uniform CV, yellow), WTMD (green), and traditional VMC (VMC, dark blue). 
        The spike in the traditional VMC error at the lowest level of $\xi(\bs)$ represents a spurious mode; see \cite{Zhang2023understanding}.}
    \label{fig:weightedvmc_cv_wtmd}
\end{figure}

Fig.~\ref{fig:weightedvmc_cv_wtmd} compares the local energy errors for traditional VMC, uniform CV sampling, and WTMD sampling.
To ensure a fair comparison, we adjusted the number of Monte Carlo samples so that all methods require the same number of wave function queries.
In the figure, 
uniform CV sampling outperforms WTMD for $s(\bs) > 0$.
In the $s(\bs) > 0$ parameter range, WTMD beats traditional VMC by a factor of $10$ to $100$ while
uniform CV sampling beats traditional VMC by a larger factor of $10^3$--$10^4$.

\begin{table}
    \centering
    \begin{tabular}{p{.25\linewidth}|p{.25\linewidth}|p{.25\linewidth}}
    Method   & Mean & Median \\
    \hline
    VMC & 2.10$\times 10^{-6}$&1.95$\times 10^{-6}$ \\ 
    Uniform CV & 1.71$\times 10^{-6}$&1.64$\times 10^{-6}$ \\ 
    WTMD & 1.59$\times 10^{-6}$ &1.52$\times 10^{-6}$\\ 
    \end{tabular} 
    
    \caption{Energy errors of traditional VMC versus weighted VMC with uniform CV sampling or WTMD, computed from ten independent trials.}
    \label{tab:energy_errorcv}
\end{table} 

In summary, the most powerful sampling distribution leverages physical insight to target important regions in the low-probability tails.
This physically-inspired VMC sampling approach also leads to competitive energy estimates, as shown in Tab.~\ref{tab:energy_errorcv}. 



\subsection{Higher-dimensional results} \label{sec:higher}

\begin{figure}
    \centering
    \includegraphics[width=1.0\columnwidth]{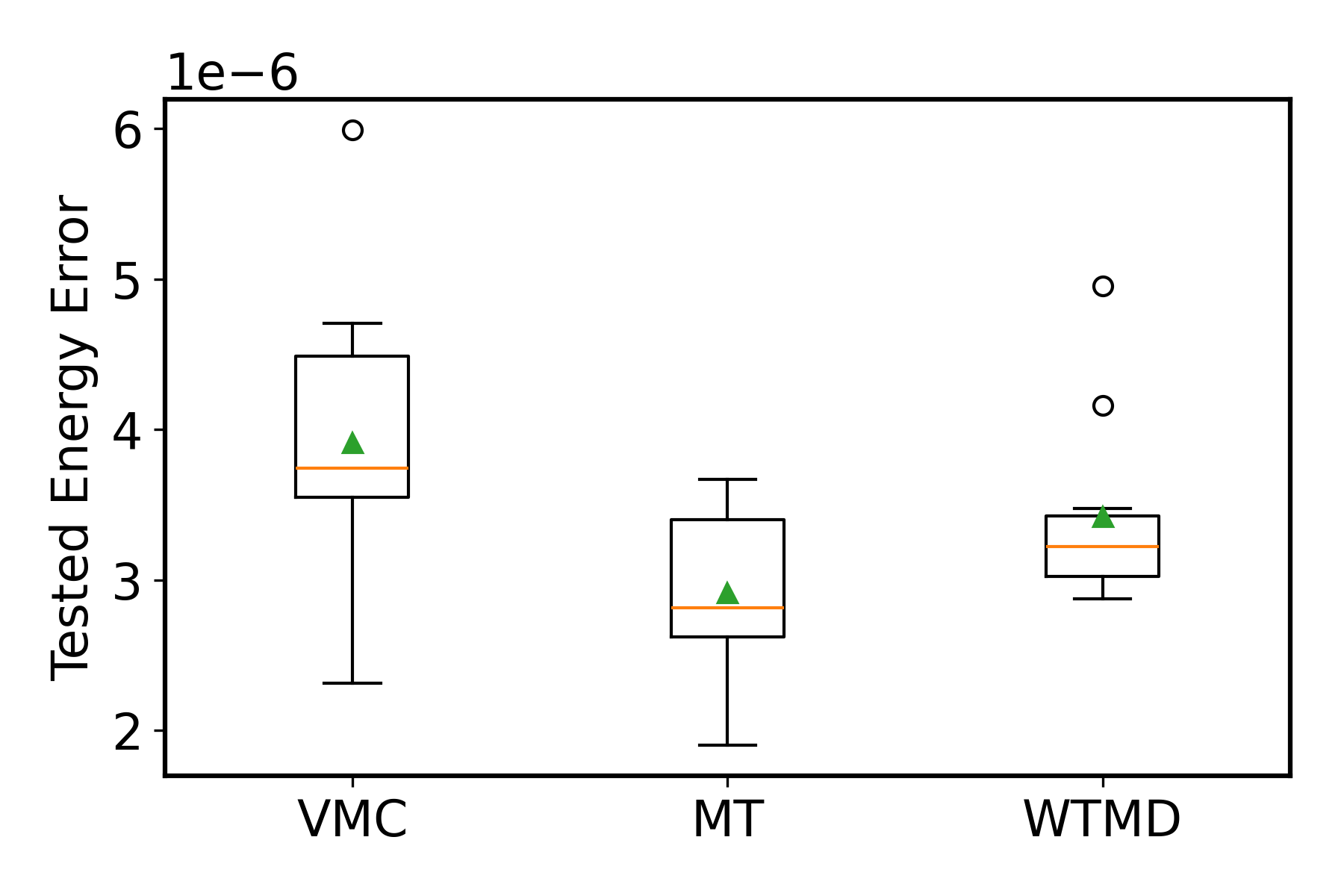}
     \includegraphics[width=1.0\columnwidth]{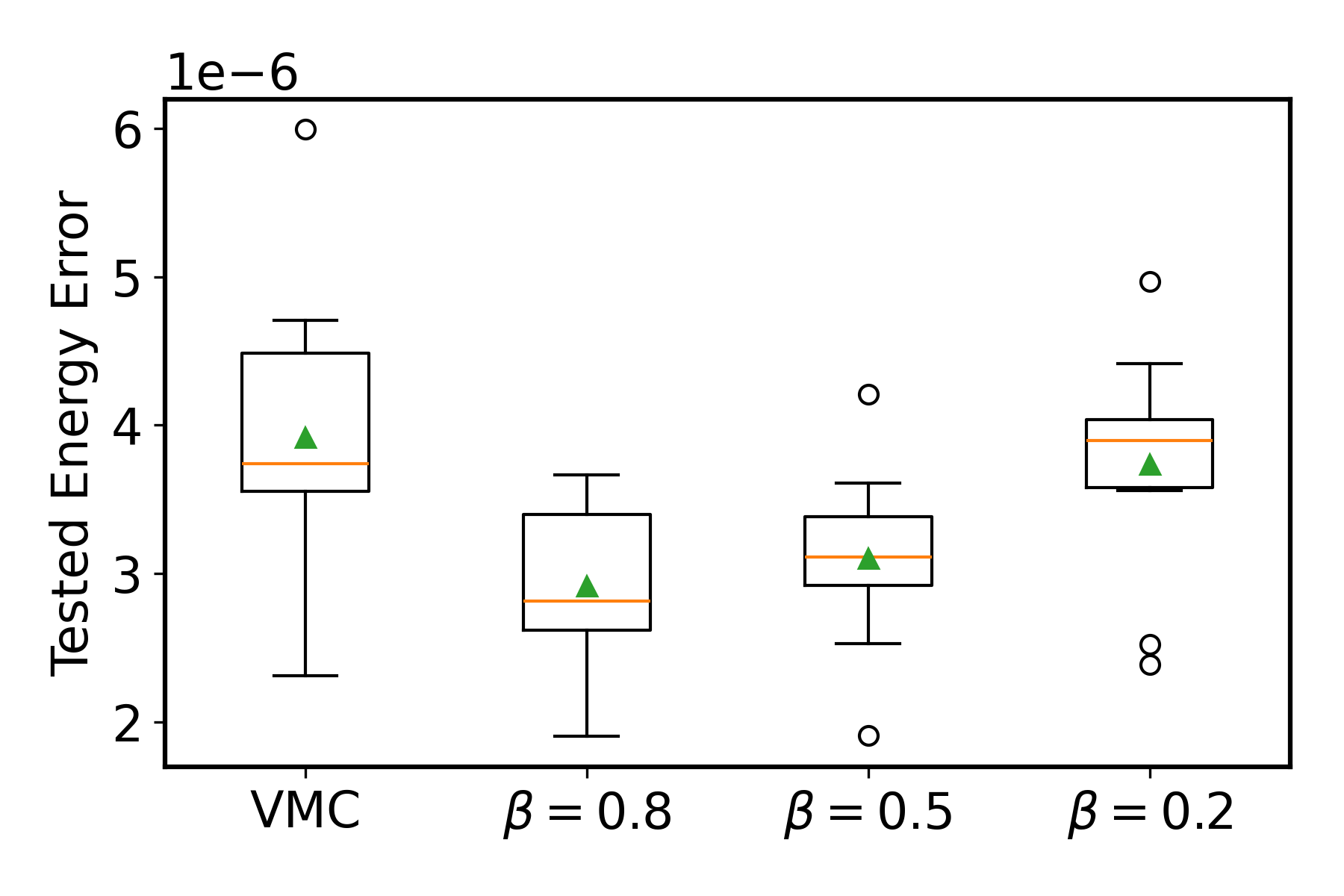}
        \caption{Boxplots of energy errors for $N = 200$ spins using traditional VMC (VMC) versus weighted VMC (MT and WTMD).
        Compare to Figs.~\ref{fig:compare_energies} and ~\ref{fig:compare_energies_temper}, which use $N = 100$ spins.}
    \label{fig:d200}
\end{figure}

Finally, we tested energy and local energy errors on a higher-dimensional XXX system with $N=200$ spins.
The results are qualitatively similar to those with $N=100$ spins, and they indicate an improvement of weighted VMC over traditional VMC.
Fig.~\ref{fig:d200} shows that MT and WTMD still outperform VMC on the higher-dimensional problem, although the WTMD results show a bit more spread than at $N=100$. 
Similar to the results in Fig.~\ref{fig:compare_energies_temper}, the energy errors from MT show a decrease-then-increase trend as the inverse temperature $\beta$ changes.
In the $N = 200$ case, the optimized parameter setting is slightly higher, $\beta=0.8$.
 
\section{Conclusion}
\label{sec:conclusion}
While neural networks are highly expressive models for quantum states, they lead to a well-known problem of generalization error. 
The learned wave functions can be inaccurate in low-probability regions, which makes them risky to use for downstream applications.

In this study, we proposed a weighted VMC method that improves the accuracy of neural network approximations in low-probability regions.
The core idea is to decouple the sampling distribution $\tilde{\rho}_\theta(\bs)$ from the wave function density $\rho_\theta(\bs)$.

We found the weighted VMC approach has several advantages over traditional VMC:
\begin{enumerate}
\item \emph{Improved energy estimation}.
Weighted VMC improves the accuracy of the estimated energies and local energies without additional computational cost. 
\item \emph{Flexible sampling}. Weighted VMC can use an arbitrary sampling distribution targeting regions where wave function refinement is needed.
\item \emph{Flexible implementation}. The method supports advanced sampling techniques such as tempering, metadynamics, and importance sampling, as well as problem-specific modifications, without any reweighting. It is broadly applicable to any quantum system with or without detailed prior knowledge of the system.
\item \emph{Extensions}. Our general approach can be extended to other optimization problems consisting of gradient flow  and projection steps. The metric used in the projection step can be designed to improve accuracy in specific regions.
\end{enumerate}
While our analysis focused on the energy and the local energy accuracy in low-probability regions, we anticipate that weighted VMC will also positively impact other observables beyond energy, such as spin correlation functions. 
\section*{Acknowledgments} 
This work was supported in part through the NYU IT High Performance Computing resources, services, and staff expertise. 
H.Z. and J.W. acknowledge support from the Army Research Office under Grant No.~W911NF-22-2-0124.

\begin{appendix}

\section{The role of width} \label{app:alpha}

\begin{table}
    \centering
    \begin{tabular}{p{.25\linewidth}|p{.25\linewidth}|p{.25\linewidth}}
        $\alpha$ & Uniform CV & VMC \\
        \hline
        6 &3.63$\times 10^{-5}$& 
        3.86$\times 10^{-3}$
        \\
        10  & 1.49$\times 10^{-5}$  & 
        1.57$\times 10^{-4}$
        \\
         20  &  4.61$\times 10^{-6}$  & 4.16$\times 10^{-6}$   \\
          30  &  2.35$\times 10^{-6}$   & 3.83$\times 10^{-6}$ \\
          50  &  2.30$\times 10^{-6}$   & 4.03$\times 10^{-6}$ 
    \end{tabular}
    \caption{Mean energy errors of traditional VMC versus uniform CV sampling, computed from ten independent trials.
    }
    \label{tab:energyalpha}
\end{table}

This appendix explores the impact of the RBM width parameter $\alpha$, which controls the total number of free parameters via $\alpha (N + 1)$.
When $\alpha$ is very low ($\alpha  \leq 10$), traditional VMC can lead to spurious modes that compromise the accuracy of the energy estimates \cite{Zhang2023understanding}.
Additionally when $\alpha$ is low ($\alpha \leq 20$), the energy estimates from weighted VMC with uniform CV sampling~\eqref{eq:mumixture} exhibit large errors.
See Tab.~\ref{tab:energyalpha}, which reports the energy errors for traditional and weighted VMC with $\alpha = 6$, $10$, $20$, $30$, $50$.

\begin{figure}
    \centering
    \includegraphics[width=1.0\columnwidth]{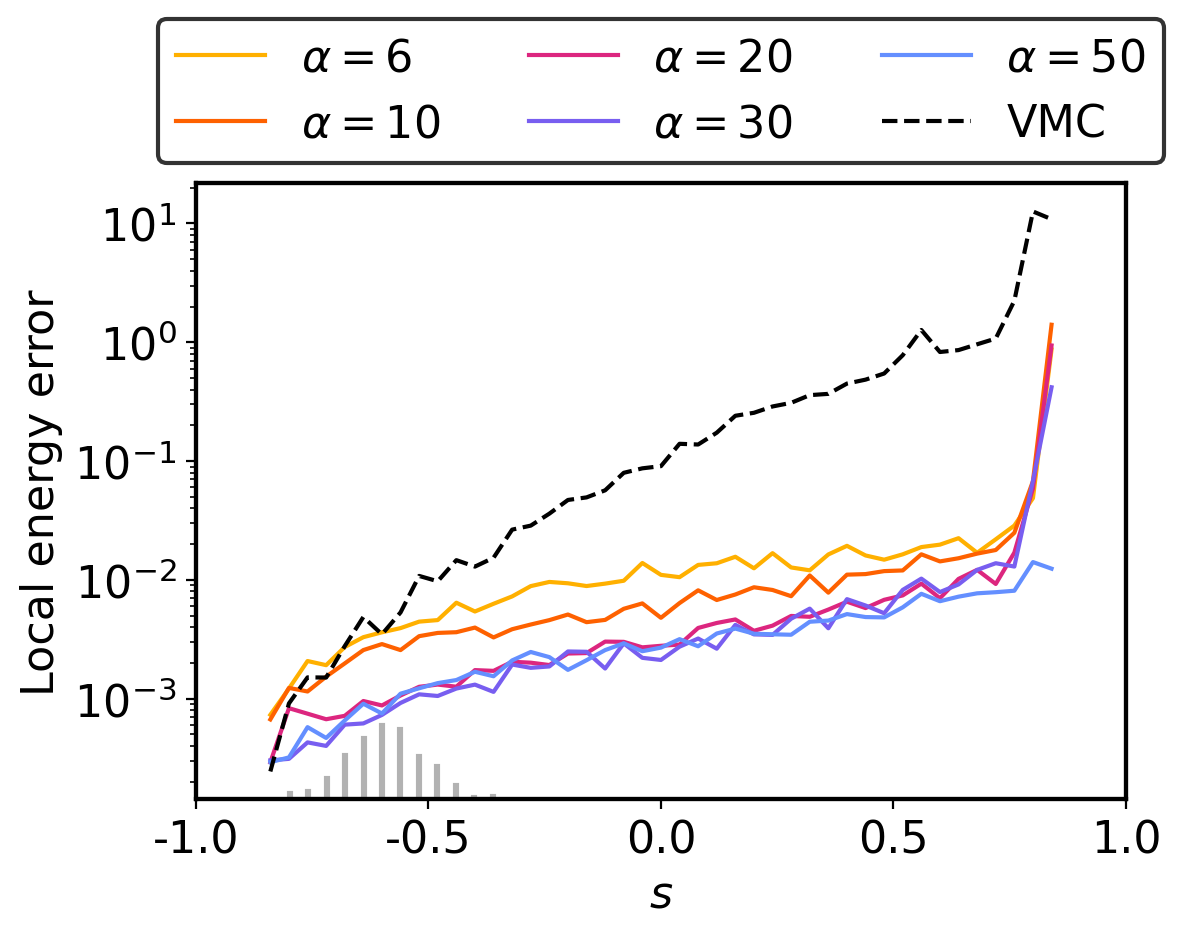} 
    \caption{Local energy errors for traditional VMC ($\alpha = 50$) versus uniform CV sampling ($\alpha = 6$, $10$, $20$, $30$, $50$).
    Histogram shows $1000$ Monte Carlo samples from $\rho_{\bt}$.
    }
    \label{fig:diff_alpha}
\end{figure}


For further insight, Fig.~\ref{fig:diff_alpha} shows the local energy errors for traditional VMC with $\alpha = 50$ versus weighted VMC with $\alpha = 6$, $10$, $20$, $30$, $50$.
When $\alpha \geq 30$, the local energy errors are uniformly smaller for weighted VMC than traditional VMC, both in the high-probability and low-probability regions.
This makes sense because the width parameter contributes flexibility to the projection step in Eq.~\eqref{eq:projection}.
Combining a flexible projection step with a weighted projection metric leads to a close match between the parametrized wave function and the gradient flow direction in both high- and low-probability regions.


\section{Details of uniform CV sampling}  
\label{app:visualize}


This appendix describes the sampling distribution $p(\bs)$ which is used for visualization in Figs.~\ref{fig:local_energy}, \ref{fig:weightedvmc_cv_wtmd}, and \ref{fig:diff_alpha} and used for sampling in the uniform CV method (Sec.~\ref{sec:uniform_cv}).
To generate samples, we introduce $L = 52$ unnormalized Gaussian density functions
\begin{equation*}
    p_i(\bs) = \exp\biggl(-\frac{(s(\bs) - m_i)^2}{2\kappa^2}\biggr)
\end{equation*}
which have the same standard deviation $\kappa = 0.04$ but different means
\begin{equation*}
    m_i = -1.06 + 0.04 i, \quad \text{for } i = 1, \ldots, 52.
\end{equation*}
To sample each density $p_i$, we initialize 10 MCMC samplers.
Then we run $2\times 10^6$ Metropolis steps and collect one sample every $2\times 10^3$ steps. 
In this way we build a pool consisting of $5.2\times 10^{5}$ samples from $p$, from which we randomly pick 1000 samples for visualization.  

\end{appendix}
\bibliographystyle{unsrt}
\bibliography{references}
\end{document}